\documentclass[aps,twocolumn,groupedaddress,showpacs]{revtex4}
\usepackage{graphicx}
\newcommand\beq{\begin{eqnarray}}
\newcommand\eeq{\end{eqnarray}}
\newcommand\bq{\begin{equation}}
\newcommand\eq{\end{equation}}
\newcommand\qom{\frac{q}{M}}
\newcommand\qove{\frac{q}{E_{\nu}}}
\newcommand\qovm{\frac{q}{2M}}
\newcommand\me{\frac{m_{\nu}}{E_{\nu}}}

\newcommand\scr{\mid C_{S}^{R} \mid^{2}}

\newcommand\invlsr{(C_{V}^{L}+ 2 M
g_{M})C_{S}^{R*}}

\newcommand\pmo{{\bf |P_{\mu}|}}

\begin{document}
\title{Scattering of neutrinos on a polarized electron target as a test
for new physics beyond the Standard Model}

\author{S. Ciechanowicz}
\email{ciechano@rose.ift.uni.wroc.pl}
\author{W. Sobk\'ow}
\email{sobkow@rose.ift.uni.wroc.pl} \affiliation{Institute of
Theoretical Physics, University of Wroc\l{}aw, Pl. M. Born 9,
\\PL-50-204~Wroc\l{}aw, Poland}
\author{M. Misiaszek}
\email{misiaszek@zefir.if.uj.edu.pl} \affiliation{ M. Smoluchowski
Institute of Physics, Jagiellonian University, ul. Reymonta 4,\\
PL-30-059 Krak\'ow, Poland }
\date{\today}
\begin{abstract}
In this paper, we analyze the scattering of the neutrino beam on
the polarized electron target, and predict the effects of two
theoretically possible scenarios beyond the Standard Model. In
both scenarios, Dirac neutrinos are assumed to be massive.

First, we consider how the existence of CP violation phase between
the complex vector V and axial A couplings of the Left-handed
neutrinos affects the azimuthal dependence of the differential
cross section. This asymmetry does not vanish in the massless
neutrino limit. The azimuthal angle $\phi_{e'}$ of outgoing
electron momentum is measured with respect to the transverse
component of the initial electron polarization {\boldmath
$\eta_{e} ^{\perp}$}. We indicate the possibility of using the
polarized electron target to measure the CP violation in the
$\nu_\mu e^-$ scattering. The future superbeam and neutrino
factory experiments will provide the unique opportunity for the
leptonic CP violation studies, if the large magnetized sampling
calorimeters with good event reconstruction capabilities are
build.

Next, we take into account a scenario with the participation of
the exotic scalar S coupling of the Right-handed neutrinos in
addition to the standard vector V and axial A couplings of the
Left-handed neutrinos.  The main goal is to show how the presence
of the R-handed neutrinos, in the above process changes the
spectrum of recoil electrons in relation to the expected Standard
Model prediction, using the current limits on the non-standard
couplings. The interference terms between the standard and exotic
couplings in the differential cross section depend on the angle
$\alpha$ between the transverse incoming neutrino polarization and
the transverse electron polarization of the target, and do not
vanish in the limit of massless neutrino. The detection of the
dependence on this angle in the energy spectrum of recoil
electrons would be a signature of the presence of the R-handed
neutrinos in the neutrino-electron scattering. To make this test
feasible, the polarized artificial neutrino source needs to be
identified.

\end{abstract}

 \maketitle

\section{\label{Sec1}Introduction}

\par Standard Model (SM) of electro-weak interactions \cite{Glashow,Wein,Salam} has a vector-axial (V-A) Lorentz structure \cite{Gell},
i.e. only left-handed (L-handed chirality states) and massless
Dirac neutrinos may take part in the charged and neutral  weak
interactions. The observed CP violation in the decays of
neutral kaons and B-mesons \cite{CP} is described by a single phase of
the CKM matrix.
\par
The vector $g_V^L$ and axial-vector $g_A^L$ neutral current coupling
constants are assumed to be real numbers, which means that $Im(g_{V}^{L}) = Im (g_{A}^{L}) = 0$.
The values of these two couplings are derived from neutrino electron scattering and
from $e^+e^- \rightarrow l^+l^-$ annihilation studies, but in the fitting procedure
the imaginary parts are fixed to their Standard Model values.
\par However, in the general case of complex $g_{V}^{L} $ and
$g_{A}^{L}$ couplings, we have one additional free parameter:
the relative phase between these couplings denoted as $\beta_{VA}$.
The CP-odd interference contribution enters the differential cross section for the scattering of left-handed neutrinos on the polarized electron target (PET), if $|\sin(\beta_{VA})| \neq 0$.
The experimental measurement of the azimuthal angle $\phi_{e'}$ of outgoing electron momentum
could be used to test the CP symmetry in lepton sector of electroweak interactions. The observation of asymmetry in the angular distribution of recoil electrons, caused by the interference terms between the standard complex couplings would give additional information about the coupling constants.
\par The magnetized sampling calorimeters (e.g. MINOS far detector) are composed of many steel leyers, which are magnetized using a coil through a hole in the center of the planes to an average field
of about 1.5 T.  In a piece of magnetized iron,  there are lots of unpaired electrons all pointing the same direction. The Moeller polarimeters determine the polarization of the electron beam by measuring
the cross section asymmetry in the scattering of polarized electrons by polarized electrons. Polarized
electrons are scattered off a polarized ferromagnetic foil, and the foil polarization is determined by measurements of the magnetization of the foil and its thickness. So, there is the well-known
and commonly used in accelerator physics technique for developing PETs.

\par Although the SM agrees well with all experimental data up
to available energies, the experimental
precision of present measurements still does not rule out the
possible participation of the exotic scalar S, tensor T and
pseudoscalar  P couplings of the right-handed (R-handed chirality
states) Dirac neutrinos beyond the  SM \cite{Delphi}.
The current  upper limits on the all non-standard couplings,
obtained from the normal and inverse muon decay, are presented in
the Table \ref{table1} \cite{Data}. In the SM, only $g_{LL}^{V}$
is non-zero value.
\begin{table}
\caption{\label{table1} Current limits on the non-standard
couplings}
\begin{ruledtabular}
\begin{tabular}{|c|c|c|}
  \hline
  Coupling constants & SM & Current limits \\
  \hline
   $|g_{LL}^V|$ & $1$ &   $>0.960$ \\
    $ |g_{LR}^V|$ & $0$ & $<0.060$ \\
    $|g_{RL}^V|$ & $0$ & $<0.110$ \\
    $|g_{RR}^V|$ & $0$ & $<0.039$ \\
    \hline
  $|g_{LL}^{S}|$ & 0 & $<0.550$\\
  $|g_{LR}^{S}|$& 0 & $<0.125$\\
  $|g_{RL}^{S}|$& 0 & $<0.424$\\
  $|g_{RR}^{S}|$& 0 & $<0.066$\\
  \hline
   $|g_{LL}^{T}|$& 0 & $0$\\
   $|g_{LR}^{T}|$& 0 & $<0.036$\\
   $|g_{RL}^{T}|$& 0 & $<0.122$\\
   $|g_{RR}^{T}|$& 0 & $0$\\
  \hline
\end{tabular}
\end{ruledtabular}
\end{table}
\par New effects due to the exotic right-handed weak interactions (ERWI)  could
be detected by the measurement of  neutrino observables (NO) which
 consist only of the interferences between the standard
${V-A}$ and ERWI, and do not depend on  the neutrino mass. The  NO
include the information on the transverse components of neutrino
spin polarization (TCNSP), both T-even and T-odd. These quantities
vanish in the SM, so the detection of the non-zero values of the
TCNSP  would be a direct signature of the R-handed neutrino
presence in the weak interactions.
The scattering of intense and polarized neutrino beam, coming from the artificial neutrino
source, on the polarized electron target could detect the effects
from the ERWI. Presently the measurement of
such observables is only theoretically possible. We give
an example how the polarized neutrino flux can be produced
if the exotic scalar coupling $C_S^R$ is present in the theory of
muon capture interaction.
\par The main goal of the first part of our paper (Section II) is to show that
the differential cross section for the $(\nu_{\mu}e^{-})$
scattering of left-handed and longitudinally polarized
muon neutrinos on the PET may be sensitive to the CP-violating effects,
if one assumes the complex standard couplings $g_{V}^{L} $,
$g_{A}^{L}$.
The main goal of the other part (Section III) is to show how the
presence of the R-handed neutrinos, in the $(\nu_{\mu}e^{-})$
scattering process, changes the energy spectrum of recoil
electrons in relation to the expected SM prediction, using the
current limits on the non-standard couplings \cite{Data}.
We concern the scattering of transversely polarized muon neutrino beam on
the PET to probe ERWI effects and include a theoretical discussion of
the possibility of developing such a beam.
\par Our analysis is model-independent and is made in the massless neutrino
limit. All the calculations are made
in the limit of vanishing neutrino mass with the Michel-Wightman
density matrices \cite{Michel} for the polarized incoming neutrino
beam (Appendix C) and for the polarized electron target,
respectively. We use the system of natural units with $\hbar=c=1$, Dirac-Pauli
representation of the $\gamma$-matrices and the $(+, -, -, -)$
metric \cite{Mandl}. \\

\section{\label{Sec2}Left handed neutrino scattering on a polarized electron target}

\par
The Standard Model (SM) of electroweak interactions is based on the gauge group
SU(2)$\times$(1). The left-handed fermion fields
$\psi_i = \left(^{\nu_i}_{l_i^-} \right) $ and $ \left( ^{u_i}_{d_i^{\prime}} \right)$
of the $i^{th}$ fermion family transform as doublets under SU(2), where
$ d_i^{\prime} \equiv \sum_j V_{ij} d_j $ and $V$ is the Cabibbo-Kobayashi-Maskawa
mixing matrix.
The vector and axial-vector couplings in SM are
\beq
g_V^L(i) & \equiv & t_3^L(i) - 2 q(i) \sin^2 \theta_W \nonumber \\
g_A^L(i) & \equiv & t_3^L(i)
\eeq
where $t_3^L(i)$ is the weak isospin of fermion $i$ ($+1/2$ for $u_i$ and $\nu_i$;
$-1/2$ for $d_i$ and $l_i$), $q_i$ is the charge of $\psi_i$ in units of $e$ and $\theta_W$ is
the weak angle. Because of the model-dependent interpretation of the coupling constants
values, they are assumed to be real numbers. For example, the total cross section for
high energy neutral-current $(\nu_\mu e^-)$ scattering is
\beq
\sigma_{SM}(\nu_\mu + e^- \rightarrow \nu_\mu + e^- ) \simeq \nonumber \\ \frac{2 G_F^2 m_e E_{\nu_\mu}}{3 \pi}
(g_V^{L2} + g_A^{L2} + g_V^L g_A^L) \, ,
\label{SMnumu}
\eeq
but in the model-independent (MI) analysis we obtain:
\beq
\sigma_{MI}(\nu_\mu + e^- \rightarrow \nu_\mu + e^- ) \simeq \nonumber \\ \frac{2 G_F^2 m_e E_{\nu_\mu}}{3 \pi}
(|g_V^L|^2 + |g_A^L|^2 + |g_V^L| |g_A^L| \cos (\beta _{VA})) \,  ,
\label{MInumu}
\eeq
where $g_V^L = |g_V^L| \,  e ^{i \beta_V^L}$, $g_A^L = |g_A^L| \,  e ^{i \beta_A^L}$ are the complex coupling constants,
$Re(g_{V}^L g_A^{L*}) = |g_V^L| |g_A^L| \cos (\beta _{VA}) $ and $\beta _{VA} = \beta _{V}^L - \beta _{A}^L$ is the relative phase between the $g_V^L$ and $g_A^L$ couplings.
\par The effective vector and axial-vector neutral coupling constants
obtained from the absolute neutrino-electron scattering event rate are
\beq
g_V^L \simeq 0 \, \, \, , \, \, \, g_A^L \simeq \pm 0.5 \, \, \, \mbox{or} \nonumber \\
g_V^L \simeq \pm 0.5 \, \, \, , \, \, \, g_A^L \simeq 0 \, \, \, .
\label{SMfit}
\eeq
However, from our MI expression (\ref{MInumu}) one can see that
the solution (with CP-violating phase):
\beq
|g_V^L| = |g_A^L| \simeq 0.35 \, \, \, \mbox{and} \, \, \, \beta_{VA} = \pm \frac{\pi}{2}
\eeq
provides to the same total cross section value as the SM fit (\ref{SMfit}).
In the next subsection we present how the existence of non zero $\beta_{VA}$
phase is related to CP-odd interference contribution in the differential cross section.
The fermion-antifermion pair production cross-sections have only T-even
contributions, but their experimental observations are essential to determine
a single soultion from possible parameters (\ref{SMfit}). Even if $\beta_{VA} = 0$
the scattering of left-handed neutrinos on the PET provides a new approach to decide
which of the two coupling types, (mainly) pure $g_A^L$ or pure $g_V^L$ coupling,
is realized in nature. This approach is model independent in contrast to $e^+e^-$ experiments
which make the assumption that the neutral current is dominated by the exchange of a
single $Z^0$.
\par As is well-known, CP violation has been observed only in the decays of
neutral kaons and B-mesons. The Standard Model describes the
existing data by a single phase of the CKM matrix,
\cite{Kobayashi}. However, the baryon asymmetry of the Universe
can not be explained by the CKM phase only, and  at least one new
source of CP violation is required \cite{barion}. The first direct
confirmation of a time reversal violation has been published by
CPLEAR Collaboration in 1998 \cite{Todd}.
 Many non-standard models take into account new CP-violating
 phases, and can be probed in  observables where the SM CP-violation is
suppressed, while alternative sources can generate a sizable
effect, e.g. the electric dipole moment of the neutron, the
transverse lepton polarization in three-body decays of charged
kaons $K^+$ \cite{Lee, Weinb}, transverse polarization of the
electrons emitted in the decay of polarized $^8Li$ nuclei
\cite{Huber}. There is no direct evidence of CP violation in the
leptonic processes, i.e.  a neutrino-electron scattering. However,
the future superbeam and neutrino factory experiments \cite{Geer}
will be able to measure the CP violating effects in the lepton
sector, where both neutrino and antineutrino oscillation will be
observed. We indicate that the scattering of neutrinos on the PET
has similar scientific possibilities.

\subsection{\label{CPVA}CP violation in standard $\nu e$ scattering }
\begin{figure*}
\begin{center}
\includegraphics[scale=.7]{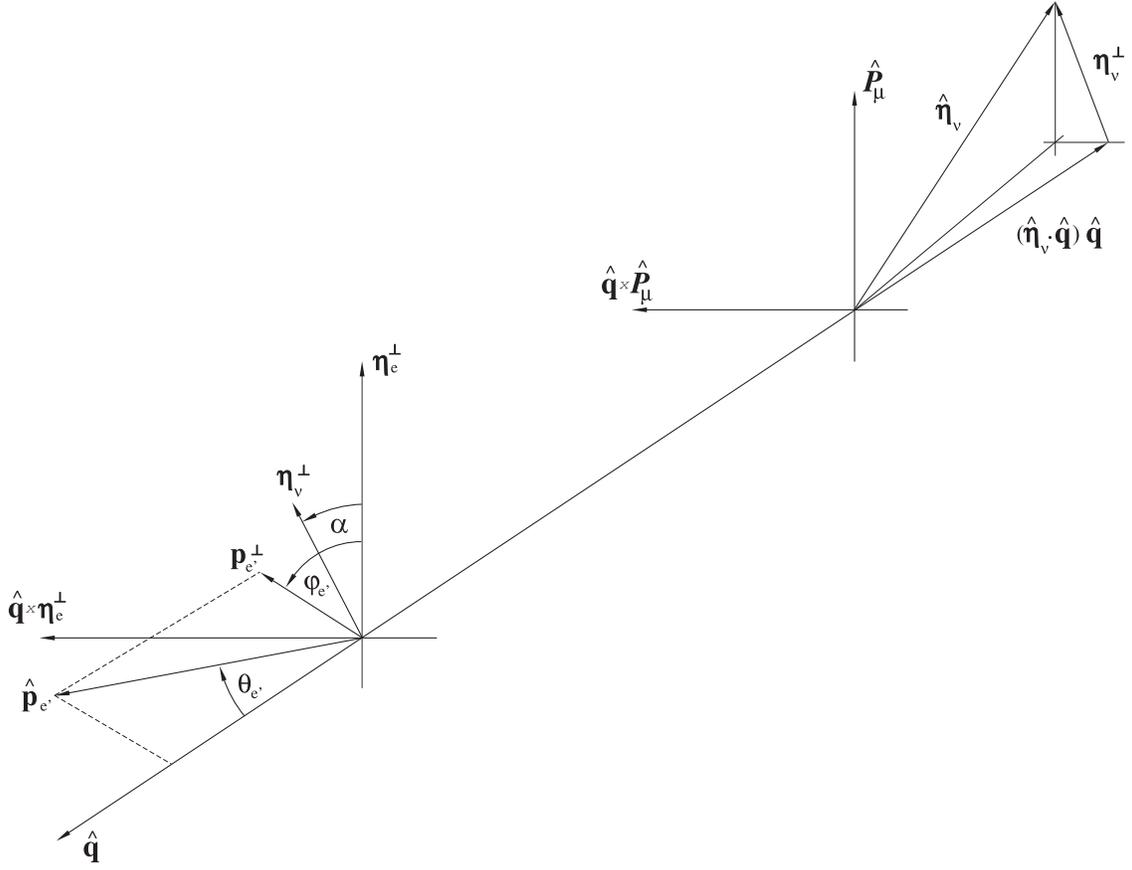}
\end{center}
\caption{Figure shows the reaction plane for the $\nu_\mu e^-$
scattering, $\mbox{\boldmath $\hat{\eta}_{\nu}$}$ - the unit
3-vector of the initial neutrino polarization in its rest frame,
$\mbox{\boldmath $\eta_{e}^\perp$}$ - the transverse electron
polarization vector of target and the production plane of
$\nu_\mu$-neutrinos for the reaction of $\mu^{-} + p \rightarrow n
+ \nu_{\mu}$. In Section II, the scattering of left handed
neutrinos is considered, which have no transverse polarization
$\mbox{\boldmath $\eta_{\nu}^\perp$} = 0$. For the considerations
described in Section III, the muon capture reaction is used as a
source of transversely polarized neutrinos. } \label{pet}
\end{figure*}

In this subsection, we consider the possibility of the CP
violation in the $\nu_\mu e^-$ scattering, when the incoming muon
neutrino beam consists only of the L-handed and
longitudinally polarized neutrinos. We assume
that these neutrinos are detected in the standard $V-A$ NC weak
interactions with the PET and both the recoil electron scattering angle
$\theta_e'$ and the azimuthal angle of outgoing electron momentum
$\phi_e'$ shown in Fig. \ref{pet} are measured with a good angular resolution.
Because we
allow for the non-conservation of the combined symmetry CP, the
amplitude includes the complex coupling constants denoted as
$g_V^L, g_A^L$ respectively to the initial neutrino of
L-chirality:
 \begin{widetext} \beq M_{\nu_{\mu} e} &=&
\frac{G_{F}}{\sqrt{2}}\Bigg\{g_{V}^{L}(\overline{u}_{e'}\gamma^{\alpha}u_{e})
(\overline{u}_{\nu_{\mu'}} \gamma_{\alpha}(1 -
\gamma_{5})u_{\nu_{\mu}})  +
 g_{A}^{L} (\overline{u}_{e'}\gamma^{5}\gamma^{\alpha}u_{e})
(\overline{u}_{\nu_{\mu'}} \gamma_{5}\gamma_{\alpha}(1 -
\gamma_{5})u_{\nu_{\mu}}) \Bigg\} , \eeq  \end{widetext}
where $ u_{e}$ and  $\overline{u}_{e'}$
$(u_{\nu_{\mu}}\;$ and $\; \overline{u}_{\nu_{\mu'}})$ are the
Dirac bispinors of the initial and final electron (neutrino)
respectively. $G_{F}= 1.16639(1)\times 10^{-5}\,\mbox{GeV}^{-2}$
\cite{Data} is the Fermi constant.

The formula for the differential cross section including the
CP-odd contribution (${\bf \hat{q}} \cdot (\mbox{\boldmath
$\hat{\eta}_{e} \times \hat{p}_{e'} $})$ is T-odd and $Im(g_{V}^L
g_A^{L*}) = |g_V^L| |g_A^L| \sin (\beta _{VA})$), proportional to
the magnitude of the transverse electron target spin polarization,
 with $\mbox{\boldmath $\hat{\eta}_{e}$} \perp {\bf \hat{ q}}$   is of the form:

\begin{widetext}
\beq \label{SMcomplex} \left(\frac{d^2 \sigma}{d y d
\phi_e'}\right)_{(V A)} & = & \frac{E_\nu
m_{e}}{4\pi^2}\frac{G_{F}^{2}}{2}(1-\mbox{\boldmath
$\hat{\eta}_{\nu}$}\cdot\hat{\bf q}) \Bigg\{ |g_{A}^{L}|^2 \left[-
\mbox{\boldmath $\hat{\eta}_{e}\cdot
\hat{p}_{e'}$}\sqrt{\frac{2m_e}{E_\nu}+y}(\sqrt{y^{3}}-2\sqrt{y})
+
\frac{m_e}{E_\nu}y + (y-2)y +2 \right] \nonumber \\
&& \mbox{} + |g_{V}^{L}|^2 \left[ y^2 - \mbox{\boldmath
$\hat{\eta}_{e}\cdot
\hat{p}_{e'}$}\sqrt{y^{3}}\sqrt{\frac{2m_e}{E_\nu}+y} -
y(\frac{m_e}{E_\nu} + 2) +2 \right]\nonumber \\
&& \mbox{} +
 Im(g_{V}^L g_A^{L*}) \, {\bf \hat{q}} \cdot
(\mbox{\boldmath $\hat{\eta}_{e} \times \hat{p}_{e'} $})\sqrt{y(\frac{2m_e}{E_\nu}+y)}\\
&& \mbox{} +
 Re(g_{V}^L g_A^{L*}) \left[
\mbox{\boldmath $\hat{\eta}_{e}\cdot
\hat{p}_{e'}$}(y-1)\sqrt{y(\frac{2m_e}{E_\nu}+y)} + (2-y)y \right]
\Bigg\} \nonumber \eeq
\end{widetext}
where $\mbox{\boldmath $\hat{\eta}_{\nu}$}\cdot\hat{\bf q} = -1 $
is the longitudinal polarization of the incoming L-handed
neutrino, ${\bf q}$ - the incoming neutrino momentum, ${\bf
p_{e'}}$ - the outgoing electron momentum, $ \mbox{\boldmath
$\hat{\eta}_{e}$}$ - the unit 3-vector of the initial electron
polarization in its rest frame, see Fig.1. The measurement of the
azimuthal angle of outgoing electron momentum $\phi_{e'}$ is only
possible when the electron target polarization is known. The
polarization vector for electrons is parallel to the magnetic
field vector. The variable $y $ is the ratio of the kinetic energy
of the recoil electron $T_{e} $ to the incoming neutrino energy
$E_{\nu} $:
\beq
y\equiv\frac{T_{e}}{E_{\nu}}=\frac{m_{e}}{E_{\nu}}\frac{2cos^{2}\theta_{e'}}
{(1+\frac{m_{e}}{E_{\nu}})^{2}-cos^{2}\theta_{e'}}. \eeq
It varies from $0 $ to $2/(2+m_e/E_\nu) $. $\theta_{e'}$ - the
polar angle between the direction of the outgoing electron
momentum  $\hat{\bf p}_{e'}$ and the direction  of the incoming
neutrino momentum $\hat{\bf q}$ (recoil electron scattering
angle), $m_{e}$ - the electron mass.
\par After the simplification of the vector products and
using the complex number identities in
the formula (\ref{SMcomplex}) for the cross section, we obtain with
$\mbox{\boldmath $\hat{\eta}_{e}$} \perp {\bf \hat{ q}} $ the new form:

\begin{figure*}
\begin{center}
\includegraphics[scale=.5]{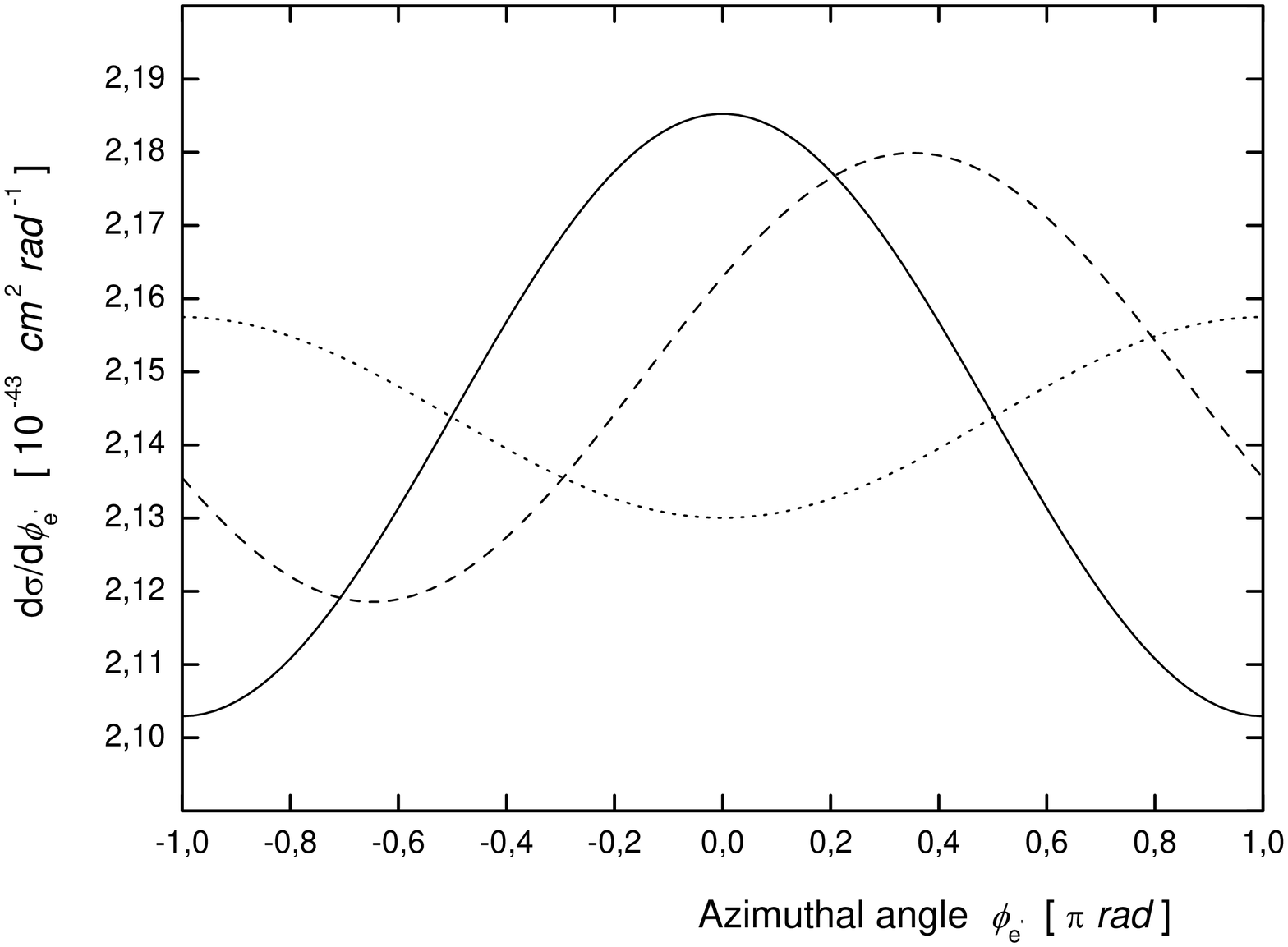}
\end{center}
\caption{Plot of the $\frac{d \sigma}{ d \phi_e'}_{(V A)}$ as a
function of the azimuthal angle $\phi_{e'}$ for the
$(\nu_{\mu}e^{-})$ scattering, $E_{\nu} = 1 \mbox{ GeV}$, $y =
0.5$, $\mbox{\boldmath$\hat{\eta}_{\nu}$}\cdot\hat{\bf q} = - 1$
and $|\mbox{\boldmath $\hat{\eta}_{e}^\perp$}| = 1$: a) the case
of the pure and real axial-vector coupling i.e. $g_A^L = 0.5$ and
$g_V^L = 0$ (solid line), b) the case of the pure and real vector
coupling i.e. $g_A^L = 0$ and $g_V^L = 0.5$ (dotted line), c) CP
violation, the case of  the complex coupling constants $|g_V^L| =
|g_A^L| = 0.354$ with the relative phase $\beta_{VA} =
\frac{\pi}{2}$ (dashed line). }  \label{VAPET}
\end{figure*}

\begin{widetext}\beq \lefteqn{\left(\frac{d^2 \sigma}{d y d \phi_e'}\right)_{(V A)} =
\frac{E_\nu m_{e}}{4\pi^2}\frac{G_{F}^{2}}{2}(1-\mbox{\boldmath
$\hat{\eta}_{\nu}$}\cdot\hat{\bf q}) \Bigg\{  |\mbox{\boldmath
$\eta_{e}^{\perp}$}|\sqrt{\frac{m_e}{E_\nu}y[2-y(2+\frac{m_e}{E_\nu})]}}\\
&& \mbox{} \cdot \bigg[ cos(\phi_{e'}) \left(2|g_V^L||g_A^L|
cos(\beta_{VA})y  + (2-y)|g_{A}^{L}|^{2} - y|g_{V}^{L}|^2 \right)
- 2|g_V^L||g_A^L|cos(\phi_{e'} + \beta_{VA}) \bigg]  \nonumber\\
&& \mbox{} +  \bigg[\left(|g_{V}^{L}|^2 + |g_{A}^{L}|^2\right)
\left( y^2 -2y +2 \right) + 2|g_V^L||g_A^L| cos(\beta_{VA})y(2-y)
- \frac{m_e}{E_\nu}y \left(|g_{V}^{L}|^2 -
|g_{A}^{L}|^{2}\right)\bigg] \Bigg\}.\nonumber \eeq
\end{widetext}

\par It can be noticed that the interference terms between the
standard $g_{V, A}^{L}$  couplings depend on the value of the $\beta_{VA}$ phase.
However, the angular asymmetry of recoil electrons is not vanishing even
if $\beta_{VA} = 0$. The CP-violating phase enters the cross section
and changes the angle at which the number of recoil electrons will be maximal ($\phi_{e'}^{max}$).

 \section{\label{Sec3} Scattering of transversely polarized muon neutrino beam
 on a polarized electron target}

\par So far the scattering of left-handed and longitudinally
polarized  neutrino beam on a polarized electron target  (SLoPET)
was proposed to probe the neutrino magnetic moments
\cite{Rashba,Trofimov} and
 the flavor composition of a (anti)neutrino beam
\cite{Minkowski}.
\par There were also the ideas of using the scattering of
transversely polarized  neutrino beam
 on the unpolarized electron target  to probe the
nonstandard properties of neutrinos. Barbieri et al. proposed to
measure the azimuthal asymmetry of recoil electrons caused by the
non-vanishing interference between the weak and electro-magnetic
interaction amplitudes \cite{Barbieri,Pastor}. Ciechanowicz et al.
\cite{Stpras} indicated that the azimuthal asymmetry of recoil
electron event rates could be generated by the interference
between the standard $(V, A)_{L}$ couplings of the L-handed Dirac
neutrinos and exotic $(S, T, P)_{R}$ couplings of the R-handed
ones in the laboratory differential cross section. All the terms
with  interference, proportional to the magnitude of the
transverse neutrino polarization, do not vanish in the massless
neutrino limit and depend on the azimuthal angle between the
transverse neutrino polarization and the outgoing electron
momentum. However, in both cases, the neutrino detectors with a
good angular resolution would have to measure both the recoil
electron scattering angle and the azimuthal angle of outgoing
electron momentum.
\par There exist the non-standard models, in which  the exotic
couplings of the Right-handed neutrinos  can appear. We mean here
three classes of such models; left-right symmetric models (LRSM),
contacts interactions (CI) and leptoquarks (LQ). For example, the
CI can be introduced both for the vector coupling of the L-handed
neutrinos  and scalar, tensor couplings of the R-handed ones,
\cite{lep}. Such interactions would allow to probe the scale for
compositeness of quarks and leptons. As to the LQ models, if the
R-handed neutrinos are taken into account, there are possible
couplings of these neutrinos to the scalar and vector LQ,
\cite{lep}.  An original discussion concerning the LQ did not
allow for the R-handed neutrinos, \cite{kwa}.
 Left-right symmetric models were proposed to explain the origin of
 the parity violation,  \cite{ Pati, Beg}. In such models
 the R-handed neutrino can couple
 to R-handed gauge boson with a mass larger than  for the observed
 standard boson $(m_1=80 GeV)$. Recently TWIST Collab. \cite{Twist} has measured the
 Michel parameter $\rho$ in the normal $\mu^+$ decay and has set new
 limit on the $W_L - W_R$ mixing angle in the LRSM. Their result $\rho= 0.75080
 \pm 0.00044 (stat.) \pm 0.00093 (syst.) \pm 0.00023$ is in good
 agreement with the SM prediction $\rho=3/4$,  and sets new upper limit
 on mixing angle $|\chi|< 0.030 \; (90 \% \; CL)$.
\par In this part of our paper we  show  that the scattering of transversely polarized muon
neutrino beam on a polarized electron target (SToPET)  may be
sensitive to the  interference effects  between the  L- and
R-handed neutrinos in the differential cross section for the
$(\nu_{\mu}e^{-})$ scattering process. Our analysis is made for
the case, when the outgoing electron direction is not observed. It
means that the azimuthal angle of the recoil electron momentum
would not be measured and nevertheless the new effects could be
observed. We consider the minimal extension of the standard $V-A$
weak interaction to indicate the new tests of the Lorentz
structure of the charged- and neutral-current weak interactions.
An admittance of all the ERWI does not change qualitatively the
conclusions from the investigations.

To show how the recoil electron spectrum  may depend on the angle
between the transverse neutrino polarization and transverse target
electron polarization, we use the muon neutrino beam produced in
the reaction where the proton at rest captures the polarized muon
$(\mu^{-} + p \rightarrow n + \nu_{\mu})$, see Appendix A. The
production plane is spanned by the direction of the initial muon
polarization ${\bf \hat{P}_{\mu}}$ (the muon is fully polarized,
i.e. $|\bf{P}_{\mu}|=1 $) and of the outgoing neutrino momentum
${\bf \hat{q}} $, Fig. \ref{pet}. ${\bf \hat{P}_{\mu}}$ and  ${\bf
\hat{q}} $ are assumed to be perpendicular to each other because
this leads to the unique conclusions as to the possible presence
of the R-handed neutrinos. When admitting additional exotic scalar
coupling $C^R_S $ in muon capture interaction, Eq. (A.1), the
outgoing muon-neutrino flux is a mixture of the L-handed neutrinos
produced in the standard $V-A $ charged weak interaction and the
R-handed ones produced in the exotic scalar $S $ charged weak
interaction (the transition amplitude and the neutrino observables
are presented in Appendix A). This mixture is detected in the
neutral current (NC) weak interaction. We mean that the incoming
L-handed neutrinos are detected in the standard $V-A $ neutral
weak interaction, while the initial R-handed ones are detected in
the exotic scalar $S $ one. Then in the final state all the
neutrinos are L-handed. Below we give the transition amplitude for
this type of neutral current: \beq \label{amp3} M_{\nu_{\mu} e}
&=&
\frac{G_{F}}{\sqrt{2}}\{(\overline{u}_{e'}\gamma^{\alpha}(g_{V}^{L}
- g_{A}^{L}\gamma_{5})u_{e}) (\overline{u}_{\nu_{\mu'}}
\gamma_{\alpha}(1 - \gamma_{5})u_{\nu_{\mu}})\nonumber \\ &  &
\mbox{} +
\frac{1}{2}g_{S}^{R}(\overline{u}_{e'}u_{e})(\overline{u}_{\nu_{\mu'}}
(1 + \gamma_{5})u_{\nu_{\mu}})\},
 \eeq
  The coupling constants are
denoted with the superscripts $L $ and $R $ as $g_{V}^{L} $,
$g_{A}^{L}$ and $g_{S}^{R}$ respectively to the incoming neutrino
of L- and R-chirality. Standard couplings $g_{V}^{L} $,
$g_{A}^{L}$ are assumed to be real, i. e. $\beta_V^L =0, \beta_A^L
=0$.

\subsection{\label{subsecA}Laboratory differential cross section}

Because we consider the case when the outgoing electron direction
is not observed,  the formula for the laboratory differential
cross section is presented after integration over the azimuthal
angle $\phi_{e'}$ of the recoil electron momentum. The result of
the calculation performed with the amplitude $M_{\nu_{\mu} e} $,
in Eq. (\ref{amp3}), is divided into three parts, standard $(V, A)
$, exotic $(S) $ and interference $(V S + A S) $: \begin{widetext}
 \beq\frac{d
\sigma}{d y } & = & \left(\frac{d \sigma}{d y }\right)_{(V, A)} +
\label{cross3} \left(\frac{d \sigma}{d y }\right)_{(S)} +
\left(\frac{d \sigma}{d y }\right)_{(V S + A S)}, \eeq
with,
\beq  \label{cross3VA} \left(\frac{d \sigma}{d y }\right)_{(V, A)}
&=& \frac{E_{\nu}m_{e}}{2\pi}\frac{G_{F}^{2}}{2}(1-\mbox{\boldmath
$\hat{\eta}_{\nu}$}\cdot\hat{\bf q})\Bigg\{\left(g_{V}^{L} +
g_{A}^{L}\right)^{2}(1 + \mbox{\boldmath
$\hat{\eta}_{e}$}\cdot{\bf\hat{q}})  + \left(g_{V}^{L} -
g_{A}^{L}\right)^{2}\left[1 - (\mbox{\boldmath
$\hat{\eta}_{e}$}\cdot {\bf \hat{q}})\left(1-
\frac{m_{e}}{E_{\nu}}\frac{y}{(1-y)}\right)\right](1-y)^{2}
\nonumber\\
& & \mbox{} - \left[\left(g_{V}^{L}\right)^{2} -
\left(g_{A}^{L}\right)^{2}\right] (1 +
\mbox{\boldmath$\hat{\eta}_{e}$}\cdot{\bf\hat{q}})\frac{m_{e}}{E_{\nu}}y\Bigg\},
\eeq
where, $\mbox{\boldmath $\eta_{\nu}$}\cdot\hat{\bf q} $ is the
longitudinal polarization of the incoming L-handed neutrino. We
shall point out that the standard (SM) part has been already
published in Ref. \cite{Rashba}. The exotic part of the cross
section may contribute merely for the R-handed neutrino scattering
($\mbox{\boldmath$\hat{\eta}_{\nu} $}\cdot\hat{\bf q}\simeq +1 $):
\beq \label{cross3S} \left(\frac{d\sigma}{dy}\right)_{(S)} &=&
\mbox{}
\frac{E_{\nu}m_{e}}{2\pi}\frac{G_{F}^{2}}{2}(1+\mbox{\boldmath$\hat{\eta}_{\nu}$}\cdot\hat{\bf
q})|g_{S}^{R}|^{2}\frac{1}{8}\left(2\frac{m_{e}}{E_{\nu}}+y
\right) y , \eeq
In the interference part we have angular correlations with the
transverse component of the neutrino polarization
{\boldmath$\eta_{\nu} ^{\perp}$}, both T-odd and T-even. The
correlation coefficients depend linearly on the exotic coupling
constant $g_{S}^{R} $. Hence, this contribution could be a tool
suitable to investigate the effects due to scalar interactions of
the R-handed neutrinos:
\beq \label{cross3VAS} \left(\frac{d \sigma}{d y }\right)_{(V S +
A S)} &=&
 - \frac{E_{\nu}m_{e}}{4\pi}\frac{G_{F}^{2}}{2} y
 \Bigg\{ \mbox{}  {\bf \hat{q}} \cdot (\mbox{\boldmath $\eta_{e}^{ \perp}$}
\times \mbox{\boldmath $\eta_{\nu} ^{\perp}$})
\mbox{}\bigg[Im(g_{V}^{L}g_{S}^{R*})\left(1+\frac{m_{e}}{2E_{\nu}}y\right)
+
Im(g_{A}^{L}g_{S}^{R*})\left(1-\frac{m_e}{2E_{\nu}}(y-4)\right)\bigg]
\nonumber\\
&&\mbox{} + (\mbox{\boldmath $\eta_{e}
^{\perp}$}\cdot\mbox{\boldmath $\eta_{\nu}
^{\perp}$})\bigg[Re(g_{V}^{L}g_{S}^{R*})
\left(1+\frac{m_{e}}{2E_{\nu}}y\right) \mbox{} +
Re(g_{A}^{L}g_{S}^{R*})\left(1-\frac{m_e}{2E_{\nu}}(y-4)\right)\bigg]\Bigg\}.
 \eeq
\end{widetext}

\par It can be noticed that the occurrence of the interference terms between the
standard $g_{V, A}^{L}$ and exotic $g_{S}^{R}$ couplings does not
depend on the neutrino mass and they pertain in the massless
neutrino limit.  The independence on the $m_\nu$ makes the
measurement of the relative phases between these couplings
possible. The terms with the interference between the standard
$g_{V, A}^{L}$ and exotic $g_{S}^{R}$ couplings, Eqs.
(\ref{cross3VAS}), include only the contributions from the
transverse component of the initial neutrino polarization
$\mbox{\boldmath $\eta_{\nu} ^{\perp}$} $ and the transverse
component of the polarized electron target $\mbox{\boldmath
$\eta_{e}^{\perp}$} $. Both transverse components are
perpendicular with respect to the $\hat{\bf q} $.
\par If one assumes the production of only L-handed neutrinos in the standard
 $(V-A)$ and  non-standard $S $ weak interactions, there is no interference
between the  $g_{V, A}^{L}$ and $g_{S}^{L} $ couplings in the
differential cross section, when $m_{\nu}\rightarrow 0$. We do not
consider this scenario.

\subsection{\label{subsecB}CP conservation in $\nu e$ scattering at low-energy}

\begin{figure*}
\begin{center}
\includegraphics[scale=.5]{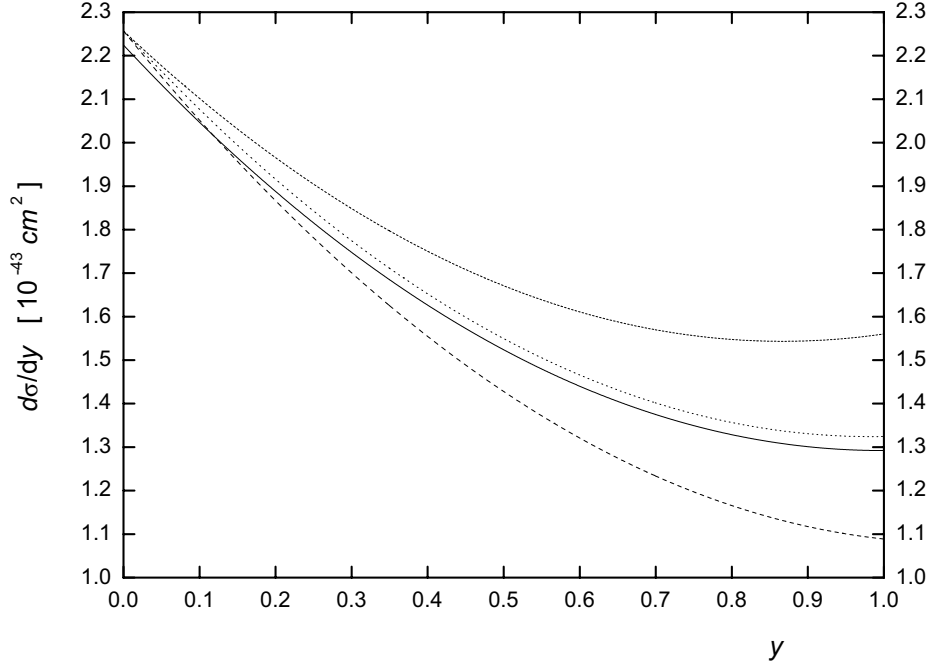}
\end{center}
\caption{Plot of the $\frac{d \sigma}{d y }$ as a function of $y$
for the $(\nu_{\mu}e^{-})$ scattering after integration over
$\phi_{e'}$, $E_{\nu}=100\, MeV$: a) SM with the L-handed neutrino
(solid line), b) CP conservation, the case of the exotic S
coupling  of the R-handed neutrinos for $\alpha=0$ (long-dashed
line), $\alpha=\pi$ (short-dashed line) and $\alpha=\pi/2, 3\pi/2$
(dotted line), respectively, c) c) CP violation, the case of the
exotic S  coupling  of the R-handed neutrinos for $\alpha +
\beta_{VS}=0$ and   $\alpha + \beta_{AS}=0$ (long-dashed line),
$\alpha + \beta_{VS}= \pi$ and   $\alpha + \beta_{AS}= \pi$
(short-dashed line), respectively.}  \label{muCP}
\end{figure*}

In this subsection, we will consider the $CP $-symmetric scenario
with the standard $(V-A)_L $ and $S_R $ weak interactions. From
the general formula for the cross section, we get with
$\mbox{\boldmath $\hat{\eta}_{e}$} \perp {\bf \hat{ q}} $ for
$\mbox{\boldmath $|\eta_{e}^{\perp}|$} = 1 $:
\begin{widetext}
 \beq \frac{d \sigma}{d y } &=& (\frac{d \sigma}{d y
})_{(V, A)} + (\frac{d \sigma}{d y })_{(S)} + (\frac{d \sigma}{dy
})_{(V S + A S)},
\\
(\frac{d \sigma}{dy })_{(V S + A S)} &=& - \label{inter3}
\frac{E_{\nu}m_{e}}{4\pi}\frac{G_{F}^{2}}{2} y |\mbox{\boldmath
$\eta_{\nu}
^{\perp}$}|\cos(\alpha)|g_S^R|\{(1+\frac{m_{e}}{2E_{\nu}}y)|g_V^L|
 + (1-\frac{m_e}{2E_{\nu}}(y-4))|g_A^L| \}, \eeq
\end{widetext}
 where $(d\sigma/d y )_{(V, A)}, (d \sigma/d y )_{(S)}$ are given by Eqs.
(\ref{cross3VA}, \ref{cross3S}) and $\alpha $ is the angle between
the $\mbox{\boldmath $\eta_{\nu} ^{\perp}$}$ and $\mbox{\boldmath
$\eta_{e}^{\perp}$} $, Fig. 1. We see that the CP-even
interference terms  enter the cross section and will be large  at
the $\alpha =0, \pi$, and they vanish for $\alpha=\pi/2, 3\pi/2$.
\par
Analyzing the assumption that the R-handed neutrinos are created
and detected in the exotic $S $ weak interaction, we have used the
same upper limit on the NC coupling $g_{S}^{R}$ as for the $\mu
$-capture coupling $C_{S}^{R} $, i.e. $|g_{S}^{R}|<0.974 $, if
allowing for the qualitative argument of weak interactions
universality. Moreover, with the presence of the exotic scalar
coupling $|g_{S}^{R}|=0.974 $, we have shifted the numerical
values of the $V $ and $A $ couplings to the new ones: $
g_{V}^{'L}= - 0.041, g_{A}^{'L}= - 0.517$, which still lie within
the experimental bars on the SM results: $g_{V}^{L}=-0.040\pm
0.015 $, $g_{A}^{L}=-0.507\pm 0.014 $ \cite{Data}, when
$\mbox{\boldmath$\hat{\eta}_{\nu}$}\cdot\hat{\bf q}= -1 $.
Independently, in Appendix A, we have estimated the transverse and
longitudinal components for the neutrino in $\mu $-capture:
$|\mbox{\boldmath $\eta_{\nu}^{\perp}$}| = 0.318 $,
$\mbox{\boldmath $\hat{\eta}_{\nu}$}\cdot\hat{\bf q}=-0.948 $.
Finally, with these estimates and the couplings $g_{V}^{'L} $,
$g_{A}^{'L}$ and $|g_{S}^{R}| $, the correlation coefficients, Eq.
(\ref{inter3}), and the cross section, Eq. (\ref{cross3}), have
been calculated in order to show the effect coming from the
R-handed muon-neutrinos.


\subsection{\label{subsecC}CP violation in $\nu e$ scattering at low-energy}

In this subsection, we analyze  the case of the violation of
combined symmetry CP. The formula for the differential cross
section including the interference contribution  between the
standard $(V-A)_L $ and  exotic $S_R $ weak interactions with
$\mbox{\boldmath $\hat{\eta}_{e}$} \perp {\bf \hat{ q}} $ for
$\mbox{\boldmath $|\eta_{e}^{\perp}|$} = 1 $  is of the form:
\begin{widetext}
\beq \frac{d \sigma}{d y } &=& (\frac{d \sigma}{d y })_{(V, A)} +
(\frac{d \sigma}{d y })_{(S)} + (\frac{d \sigma}{d
y })_{(V S + A S)},\\
(\frac{d \sigma}{d y })_{(V S + A S)} & = & - \label{inter3V}
\frac{E_{\nu}m_{e}}{4\pi}\frac{G_{F}^{2}}{2} y
 |\mbox{\boldmath $\eta_{\nu}
^{\perp}$}||g_S^R|\{(1+\frac{m_{e}}{2E_{\nu}}y)|g_V^L|\cos(\alpha
+ \beta_{V S}) + (1-\frac{m_e}{2E_{\nu}}(y-4))|g_A^L|\cos(\alpha +
\beta_{A S})\}, \eeq \end{widetext}
where $\beta_{VS} \equiv \beta_{V}^{L} - \beta_{S}^{R}$, $\beta_{A
S} \equiv \beta_{A}^{L} - \beta_{S}^{R} $ are  the relative phases
between the $g_{V}^{L}$, $g_{S}^{R}$ and $g_{A}^{L}$, $g_{S}^{R}$
couplings, respectively.
\par It can be seen that the CP-odd interference contribution enters
the cross section and will be substantial at  the $\alpha +
\beta_{V S}=0,\pi$ and $\alpha+\beta_{A S} = 0, \pi,$ and it
vanishes for the $\alpha + \beta_{V S}=\pi/2$ and $\alpha+\beta_{A
S} = \pi/2$, respectively. The situation is illustrated in the
Fig. \ref{muCP}  for the same limits as for the CP-symmetric case
with $E_{\nu}=100 \, MeV$ (long-dashed and short-dashed lines,
respectively). The phases $\beta_{VS} $ and $\beta_{AS} $ in Eq.
(\ref{inter3V}), when different from $0 $ or $\pi $, may result
from $CP $-violation in NC weak interaction ($\nu_\mu e^- $). The
angle $\alpha $ is defined in accordance with Fig. 1 and relates
the direction of $\mbox{\boldmath $\eta$}_{\nu}^{\perp} $ to the
direction of $\mbox{\boldmath $\eta$}_{e}^{\perp} $. So with the
proper choices of $\alpha $, the phases $\beta_{VS} $ and
$\beta_{AS} $ could be detected by measuring the maximal asymmetry
of the cross section $d\sigma/dy $. \par On the other hand, if
knowing these phases prior to ($\nu_\mu e^- $) scattering, it
would be possible to test $CP $-symmetry in muon capture. In case
of $CP $-violation, the neutrino transverse polarization vector
$\mbox{\boldmath $\eta$}_{\nu}^{\perp}$ would be turned aside from
the production plane $(\hat{\bf q},\hat{\bf P}_\mu ) $, having the
$CP $-breaking component
${\bf <S_{\nu}\cdot({\hat P}_{\mu}\times\hat{q})>}_{f} $, see Fig.
1 and Eq. (A3).
For illustration purposes let us take $\beta_{VS}=\beta_{AS}=0 $
(i.e. $CP $-symmetry in $\nu_\mu e^- $). Next, we measure the
angular correlation
$\mbox{\boldmath $\eta$}_{e}^{\perp}\cdot\mbox{\boldmath
$\eta$}_{\nu}^{\perp} \sim\cos(\alpha) $,
in order to see the direction of $\mbox{\boldmath
$\eta$}_e^{\perp} $ along which the maximal asymmetry is oriented.
If this direction were turned aside from $\hat{\bf P}_\mu $, it
would be evidence for $CP $-breaking.

\section{Conclusions}

In the first part of the paper, we show that  the SLoPET can be
used to measure the CP violation in the pure leptonic process,
Fig. 2. The azimuthal asymmetry of the recoil electrons does not
depend on the neutrino mass and is not vanishing even if
$\beta_{VA} = 0$. The CP-breaking phase $\beta_{VA}$ could be
detected by measuring the maximal asymmetry of the cross section.
The future superbeam and neutrino factory experiments will provide
the unique opportunity for the leptonic CP violation studies, if
the large magnetized sampling calorimeters with good event
reconstruction capabilities are build.
\par In the second part, we show that the SToPET  may
be used to detect the effects caused by the interfering L- and
R-handed neutrinos. In spite of the integration over the azimuthal
angle $\phi_{e'}$ of the recoil electron momentum, the  terms with
the interference between the standard $(V, A)_{L}$ and exotic
$S_{R}$ couplings in the laboratory differential cross section
depend on the angle $\alpha$, Fig. 1, between the transverse
incoming neutrino polarization and  the transverse electron
polarization of the target and are present even in  the massless
neutrino limit. The observation of the dependence on this angle
$\alpha$ in the recoil electron energy spectrum  would be a clear
signal of the R-handed neutrinos in the $\nu e$ scattering.
\par
It can be noticed that
the disagreement with the SM would be substantial for the small
polar angle $\theta_{e'}$, both for the CP-even and CP-odd cases,
Fig. 3.
\par To search for the effects connected with the ERWI,  the strong  polarized neutrino beam
and the polarized electron target is required. The electron target
should be polarized perpendicular to the direction of the incoming
neutrino beam, $\mbox{\boldmath $\hat{\eta}_{e}$}\cdot {\bf
\hat{q}}=0$, because it leads to the unique conclusions as to the
R-handed neutrinos. If one has the polarized artificial neutrino
source, the direction of the transverse neutrino polarization with
respect to the production plane will be fixed. So having the
assigned direction of the  polarization axis  of electron target
and turning the polarization axis of neutrino source, the
dependence of the event number  on the angle $\alpha$ could be
tested.
\par
It seems worthy of exploring high energy region in the $L $- and
$R $- handed neutrinos interference. Because of the angular
correlation between the transverse spin polarizations of the
neutrino and electron and at the large kinetic energy transfer to
the recoil electron, we see strong angular asymmetry in the cross
section $d\sigma/dy $ for the small values of the polar recoil
angle, see Eq. (\ref{inter3}, \ref{inter3V}). We expect for this
fact some interest in the accelerator laboratories working with
neutrino beams, accompanied by the progress in the spin
polarization engineering. For the future outline, we shall inspect
the other examples, which could be interesting from the point of
observable
 effects caused by the exotic neutrino states. We plan to work mainly
on the weak interaction processes that are known from experiment
or have been already under consideration in the literature.


\begin{acknowledgments}
This work was supported in part by the grant 2P03B 15522 of The
Polish Committee for Scientific Research and by The Foundation for
Polish Science.
\end{acknowledgments}
\begin{widetext}
 \appendix
\section{ Muon capture by proton}
The amplitude for the muon capture by proton $(\mu^{-} + p
\rightarrow n + \nu_{\mu})$, as a production process of massive
muon-neutrinos, is commonly of the form: \beq M_{\mu^-} & = &
(C_{V}^{L}+ 2 M g_{M})({\overline u}_{\nu}
    \gamma_{\lambda}(1 - \gamma_{5})u_{\mu})
    ({\overline u}_{n}\gamma^{\lambda}u_{p})
+ (C_{A}^{L}+
  m_{\mu}\qovm g_{P})({\overline u}_{\nu}
    i\gamma_{5}\gamma_{\lambda}(1 - \gamma_{5})u_{\mu})
    ({\overline u}_{n}i\gamma^{5}\gamma^{\lambda}u_{p})\nonumber\\
&& \mbox{} + C_{S}^{R}({\overline u}_{\nu}(1 -
\gamma_{5})u_{\mu})({\overline u}_{n} u_{p}),
   \eeq
where the fundamental coupling constants are denoted as $C_{V}^{L}
$, $C_{A}^{L}$ and $C_{S}^{R} $ respectively to the outgoing
neutrino of L- and R-chirality. Since we do not preclude the $CP
$-asymmetry between SM and exotic sectors, we allow for these
coupling constants to be the complex numbers. $g_{M}, g_{P}$ - the
induced weak couplings of the left-handed neutrinos, i.e. the weak
magnetism and induced pseudoscalar, respectively;  $m_{\mu} $, $q
$, $E_{\nu} $, $m_{\nu} $, $M $ - the muon mass, the absolute
value of the neutrino momentum, its energy, its mass and the
nucleon mass; $u_{p} $, ${\overline u}_{n} $ - the Dirac bispinors
of initial proton and final neutron; $u_{\mu} $, ${\overline
u}_{\nu}$ - the Dirac bispinors of initial muon and final
neutrino.
\par Following the results of Ref. \cite{Sobkow}, in the case of
non-vanishing neutrino mass $(m_{\nu}\not =0)$, we take the
transverse components of the neutrino spin polarization, T-even:
\beq \lefteqn{{\bf <S_{\nu}\cdot{\hat P}_{\mu}>}_{f}}\nonumber\\
  & =  & \pmo \{(1 + \qove\qovm)Re(\invlsr)
 + \frac{1}{2} \me( |C_{V}^{L} + 2 M g_{M}|^{2} - |C_{A}^{L} +
  m_{\mu}\qovm  g_{P}|^{2} + \scr)\},
\eeq and T-odd: \beq {\bf <S_{\nu}\cdot({\hat
P}_{\mu}\times\hat{q})>}_{f} &=& - \pmo (\qove +
\qovm)Im(\invlsr).
\eeq Here, ${\bf S_{\nu}}$ is  the  neutrino spin operator, $s=1/2
$ is the neutrino spin; ${\bf \hat{P}_\mu} $ is the unit vector of
the muon polarization in the muonic atom $1s $ state, and $\bf\hat
q $ is the unit vector of the neutrino momentum. ${\bf
\hat{P}_\mu} $ and $\bf\hat q $ are perpendicular to each other,
${\bf \hat{P}_\mu \cdot  \hat{q}} = 0 $. \par It can be noticed
that in the limit of vanishing neutrino  mass, these observables
consist only of the interference term between the standard
$C_{V}^{L}$ coupling and exotic  $C_{S}^{R}$ one. There is no
contribution to these observables from the SM in which neutrinos
are only L-handed and massless. The neutrino mass terms, $m_\nu
/E_\nu $, in the above observables give  a very small contribution
in relation to the main one coming from the interference terms and
they are neglected in the considerations.
As the induced weak couplings enter additively  the fundamental
$C_{V,A}^{L} $ couplings, they are  omitted in the considerations,
for their presence does not change qualitatively the conclusions
concerning the transverse neutrino polarization.
%
\par Using the current data \cite{Data}, we calculate the lower limits on the SM
couplings: $|C_{V}^{L}|>0.850 (4G_{F}/\sqrt{2})\cos\theta_{c} $
and $|C_{A}^{L}|>1.070 (4G_{F}/\sqrt{2})\cos\theta_{c} $, and
upper limit on the exotic scalar: $|C_{S}^{R}|<0.974
(4G_{F}/\sqrt{2})\cos\theta_{c} $. Now, we may give the upper
bound on the magnitude of the transverse neutrino polarization in
the massless neutrino limit:
\beq |\mbox{\boldmath $\eta_{\nu }^{\perp}$}| & = & {1\over
s}\sqrt{{\bf < S_{\nu}\cdot({\hat P}_{\mu}\times {\hat q})>}^{2} +
{\bf <S_{\nu}\cdot {\hat P}_{\mu}>}^{2}}. \eeq
The transverse components, Eqs. (A.2) and  (A.3), are calculated
with the amplitude $M_{\mu^-} $ and normalized with the $\mu
$-capture probability ${\bf <1>_{f}} $:
\beq |\mbox{\boldmath $\eta_{\nu }^{\perp}$}| = {1\over
s}\left|\frac{C_{S}^{R}}{C_{V}^{L}}\right|\left(1+\qovm\right)
\left[1+\qom +
\left(3+\qom\right)\left|\frac{C_{A}^{L}}{C_{V}^{L}}\right|^{2} +
 \mbox{} + \left|\frac{C_{S}^{R}}{C_{V}^{L}}\right|^{2} -
2\qom\left|\frac{C_{A}^{L}}{C_{V}^{L}}\right|
\cos(\alpha_{AV}^{L})\right]^{-1}.  \eeq
After inserting from above the limits on coupling constants and
with the relative phase between the standard $C_{A}^{L}$ and
$C_{V}^{L}$ couplings $\alpha_{AV}^{L}\equiv \alpha_{A}^{L} -
\alpha_{V}^{L} = \pi $, under the condition that $|\mbox{\boldmath
$\hat{\eta}_{\nu}$}|=1 $, one obtains; $|\mbox{\boldmath
$\eta_{\nu}^{\perp}$}| \leq 0.318 $, which means that the value of
the longitudinal neutrino polarization is equal to
$\mbox{\boldmath $\hat{\eta}_{\nu}$}\cdot\hat{\bf q} = -0.948 $.

\section{ Four-vector neutrino
polarization and Michel-Wightman density matrix} The formulas for
the 4-vector of the massive neutrino polarization $S $ in its rest
frame and for the initial neutrino moving  with the momentum ${\bf
q}$, respectively, are as follows:
\beq S & = & (0,\mbox{\boldmath $\hat{\eta}_{\nu}$}),\\
 S' & = & \frac{\mbox{\boldmath $\hat{\eta}_{\nu}$}\cdot{\bf q}}{E_{\nu}}\cdot
\frac{1}{m_{\nu}} \left(
\begin{array}{c}  E_{\nu}\\ {\bf q} \end{array} \right)
 + \left(
\begin{array}{c}  0\\ \mbox{\boldmath $\hat{\eta}_{\nu}$}  \end{array} \right) -
\frac{\mbox{\boldmath $\hat{\eta}_{\nu}$}\cdot{\bf
q}}{E_{\nu}(E_{\nu}+m_{\nu})} \left( \begin{array}{c}  0\\ {\bf q}
\end{array} \right), \\
S^{0'} & = & \frac{{|\bf q|}}{m_{\nu}}(\mbox{\boldmath
$\hat{\eta}_{\nu}$}\cdot{\bf \hat{q}}), \\
{\bf S'} & = & \frac{E_{\nu}}{m_{\nu}}(\mbox{\boldmath
$\hat{\eta}_{\nu}$}\cdot{\bf \hat{q}}){\bf \hat{q}} +
\mbox{\boldmath $\hat{\eta}_{\nu}$} - (\mbox{\boldmath
$\hat{\eta}_{\nu}$}\cdot{\bf \hat{q}}){\bf \hat{q}},
 \eeq
 where $\mbox{\boldmath $\hat{\eta}_{\nu}$}$ - the unit vector of the
 initial neutrino polarization in its rest frame.
 The formula for the Michel-Wightman density matrix \cite{Michel} is given by:
 \beq
\Lambda_{\nu}^{(s)}  &=& \mbox{} \sum_{r=1,
2}u_{r}\overline{u}_{r} \sim
 [(q^{\mu}\gamma_{\mu}) + m_{\nu} +
\gamma_{5}(S^{'\mu}\gamma_{\mu})(q^{\mu}\gamma_{\mu}) +
\gamma_{5}(S^{'\mu}\gamma_{\mu}) m_{\nu}],  \\
(S^{'\mu}\gamma_{\mu}) &=& \frac{\mbox{\boldmath
$\hat{\eta}_{\nu}$}\cdot{\bf q}}{E_{\nu}m_{\nu}}(q^{\mu}
\gamma_{\mu}) - (\mbox{\boldmath $\hat{\eta}_{\nu}$} -
\frac{(\mbox{\boldmath $\hat{\eta}_{\nu}$}\cdot{\bf q}){\bf
q}}{E_{\nu}(E_{\nu}+
m_{\nu})})\cdot\mbox{\boldmath$\gamma$}, \\
(S^{'\mu}\gamma_{\mu})(q^{\mu}\gamma_{\mu}) &=&
\frac{m_{\nu}}{E_{\nu}}\mbox{\boldmath
$\hat{\eta}_{\nu}$}\cdot{\bf q} - (\mbox{\boldmath
$\hat{\eta}_{\nu}$} - \frac{(\mbox{\boldmath
$\hat{\eta}_{\nu}$}\cdot{\bf q}){\bf q}}{E_{\nu}(E_{\nu}+
m_{\nu})})\cdot \mbox{\boldmath $\gamma$}(q^{\mu}\gamma_{\mu}),
\\ (S^{'\mu}\gamma_{\mu}) m_{\nu} &=& \frac{\mbox{\boldmath
$\hat{\eta}_{\nu}$}\cdot{\bf q}}{E_{\nu}}(q^{\mu}\gamma_{\mu}) -
m_{\nu}(\mbox{\boldmath $\hat{\eta}_{\nu}$} -
\frac{(\mbox{\boldmath $\hat{\eta}_{\nu}$}\cdot{\bf q}){\bf
q}}{E_{\nu}(E_{\nu}+ m_{\nu})})\cdot \mbox{\boldmath $\gamma$},
\eeq and in the limit of vanishing  neutrino mass $m_{\nu} $, we
have \beq \lim_{m_{\nu}\rightarrow 0} \Lambda_{\nu}^{(s)}
 & = & \mbox{} \left[1 + \gamma_{5}\left(\frac{\mbox{\boldmath
$\hat{\eta}_{\nu}$} \cdot{\bf q}}{|{\bf q}|} - (\mbox{\boldmath
$\hat{\eta}_{\nu}$} - \frac{(\mbox{\boldmath
$\hat{\eta}_{\nu}$}\cdot{\bf q}){\bf q}}{|{\bf q}|^{2}})\cdot
\mbox{\boldmath
$\gamma$}\right)\right]\left(q^{\mu}\gamma_{\mu}\right). \eeq
\end{widetext}
 We
see that in spite of the singularities $m_{\nu}^{-1}$ in the
polarization four-vector $S^\prime $, the density matrix
$\Lambda_{\nu}^{(s)}$ remains finite including the transverse
component of the neutrino spin polarization.

\end{document}